\documentclass[twocolumn, superscriptaddress, nofootinbib]{revtex4-1}

\usepackage[english]{babel}
\usepackage[pdftex]{graphicx}
\usepackage{color}
\usepackage{float}

\usepackage[sort&compress]{natbib}
\usepackage{hyperref}
\usepackage{amsmath}
\usepackage{amssymb}
\usepackage{mathtools}
\usepackage{braket}

\renewcommand{\vec}[1]{\ensuremath{\boldsymbol{#1}}}
\newcommand{\un}[1]{\ensuremath{\,\mathrm{#1}}}
\newcommand{\fig}[1]{Figure~\ref{fig:#1}}
\newcommand{\eq}[1]{(\ref{#1})}

\newcommand{\I}{\mathrm{i}}
\newcommand{\abs}[1]{\left| #1 \right|}

\usepackage{lineno}

\begin{document}

\title{Phosphorene pnp junctions as perfect electron waveguides}

\author{Yonatan~Betancur-Ocampo}
\email{ybetancur@icf.unam.mx}
\affiliation{Instituto de Ciencias F\'isicas, Universidad Nacional Aut\'onoma de
  M\'exico, Cuernavaca 62210, M\'exico}

\author{Emmanuel Paredes-Rocha}
\affiliation{Instituto de Ciencias F\'isicas, Universidad Nacional Aut\'onoma de
  M\'exico, Cuernavaca 62210, M\'exico}

\author{Thomas~Stegmann}
\email{stegmann@icf.unam.mx}
\affiliation{Instituto de Ciencias F\'isicas, Universidad Nacional Aut\'onoma de
  M\'exico, Cuernavaca 62210, M\'exico}

\begin{abstract}
  The current flow in phosphorene pnp junctions is studied. At the interfaces of the junction,
  omni-directional total reflection takes place, named anti-super-Klein tunneling, as this effect is
  not due to an energetically forbidden region but due to pseudo-spin blocking. The anti-super-Klein
  tunneling confines electrons within the junction, which thus represents a perfect lossless
  electron waveguide. Calculating the current flow by applying the Green's function method onto a
  tight-binding model of phosphorene, it is observed that narrow electron beams propagate in these
  waveguides like light beams in optical fibers. The perfect guiding is found for all steering
  angles of the electron beam as the total reflection does not rely on the existence of a critical
  angle. For low electron energies and narrow junctions, the guided modes of the waveguide are
  observed. The waveguide operates without any loss only for a specific orientation of the
  junction. For arbitrary orientations, minor leakage currents are found, which however decay for
  low electron energies and grazing incidence angles. It is shown that a crossroad shaped pnp
  junction can be used to split and direct the current flow in phosphorene. The proposed device, a
  phosphorene pnp junction as a lossless electron waveguide, may not only find applications in
  nanoelectronics but also in quantum information technology.
\end{abstract}

\maketitle

\section{Introduction}

Graphene is nowadays certainly the most studied two-dimensional (2D) material, due to its
exceptional properties and possible technological applications. Nevertheless, the absence of an
intrinsic band gap complicates the realization of certain nanoelectronic devices
\cite{Neto2009}. Several other 2D materials have been predicted and synthesized in recent years
\cite{Naumis, Das2015, Bhimanapati2015, Geng2018, Wehling2014, Li2014, Vogt2012, Li2018b,
  Li2014b}. One of them is phosphorene, a single layer of black phosphorous, which possesses an
intrinsic band gap and high electron mobility \cite{Das2014, Na2014, Li2014, Li2014a, Koenig, Ezawa,
  Peng, Rudenko, Cakir, Lv, Chang2015, Wu2015, Kou, Lew, Sisakht, Rudenko2, Elahi, Utt, Ameen2016,
  Li2, Carvalho, Mehboudi2016, Liu2017, Soleimanikahnoj2017, Yang2017,Lewenkopf, Sarkar, Quhe2018,
  Zhu, Partoens, Quhe2018, Wu2018, Zhu, Zhang, Betancur-Ocampo2019, Miao, Watts, Ray, Jung2020,
  Dana2020}. Moreover, unconventional properties like negative reflection and anti-super-Klein
tunneling \cite{Betancur-Ocampo2019}, make phosphorene a promising material for nanoelectronic
applications.

The ballistic electron propagation in 2D materials allows the observation of analogies from optics
\cite{Betancur-Ocampo2019, Betancur, Betancur3, Stegmann2016, Stegmann2019, Fuchs, Chen, Pozzi,
  Elahi2019, Cheianov, Cheianov2, Lee, Baeuerle2018, Li2018a, Sajjad2011, Bai2018, Graef2019,
  Zhang2019, Yang2019}. For example, it has been shown that an electron beam, which hits the
interface of a pn junction, is refracted similar to a light beam at the interface of two different
media. An important tool for electron optics in 2D materials are waveguides or fibers, which may
find application as electron conveyors in quantum information processes \cite{Baeuerle2018}. One
strategy to construct such waveguides is to use gates that generate regions of different
electrostatic potential and thus, constitute a pnp junction or, in other words, a potential well.

Electron optics in 2D materials is studied commonly in graphene. However, the Klein tunneling of
electrons prevents the efficient confinement of electrons by means of pnp junctions
\cite{Katsnelson, Kim, Rozhkov2011}. As a solution to this problem, it has been suggested to steer
the electron beam with incidence angles larger than the critical angle or to use smooth bipolar
junctions \cite{Wu2011,Zhang2009, Hartmann2010, Yuan2011, Myoung2011, Huang2012, Beenakker2009,
  Zhao2010, Williams2011, Rickhaus2013, Rickhaus, Li2018a, Wilmart2014, Rickhaus2015, Cupo, Li2018,
  Lee2018}. Nevertheless, these strategies do not prevent Klein tunneling at normal incidence and
hence, reduce the efficiency of the device. In this paper, we show that these problems do not appear
in phosphorene due to anti-super-Klein tunneling, the omni-directional total reflection at the
interface of a pn junction.

In the following, we analyze quantum transport in phosphorene pnp junctions. We calculate
numerically the current flow by applying the nonequilibrium Green's function method on a
tight-binding model of phosphorene. These calculations demonstrate that phosphorene pnp junctions
constitute perfect electron waveguides, without any leakage current through the sidewalls. This
perfect guiding of electrons persists for all energies in the pnp regime and all incidence
angles. The electron waveguide works perfectly only for a specific orientation of the junction with
respect to the phosphorene lattice. For arbitrary orientations, minor leakage currents are observed,
which however decay for low electron energies and grazing incidence angles. Using a continuum model
of phosphorene, we solve the Schr\"odinger equation analytically, which will allow us to understand
the appearance of guided modes for low electron energies and narrow waveguides. Finally, we
demonstrate that the current can be divided and guided efficiently in a crossroad-shaped pnp
junction.

\section{Tight-binding model of phosphorene}

\begin{figure*}[t]
  \centering
  \includegraphics[scale= 0.37]{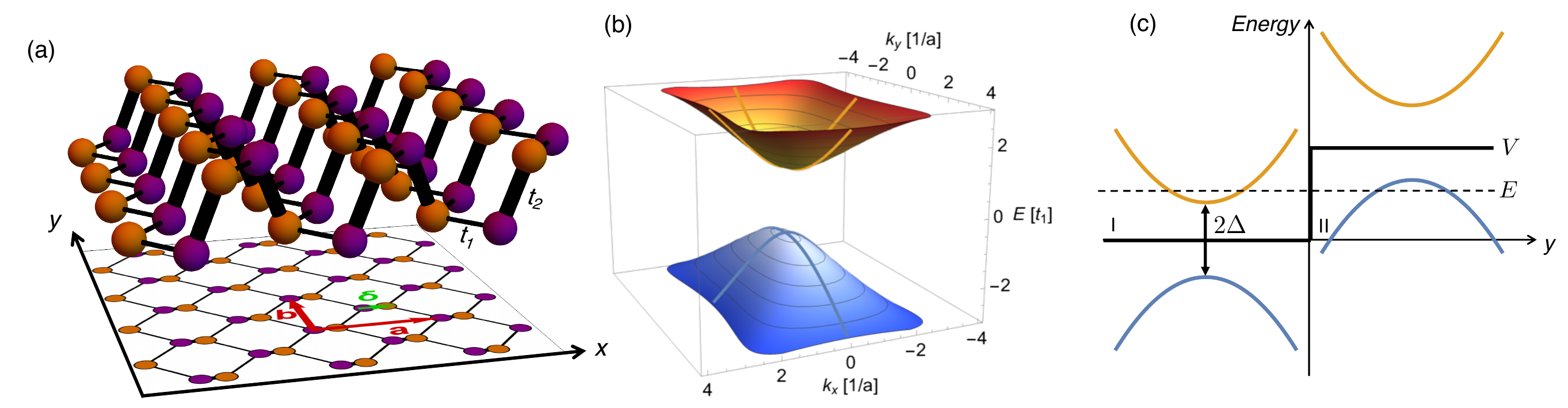}
  \caption{(a) Crystal structure of phosphorene and its projection on the $xy$ plane. The parameters
    $t_1$ and $t_2 \approx -3t_1$ represent the hoppings to first and second nearest neighbors,
    while $\vec{a}$ and $\vec{b}$ are the lattice vectors. The vector $\vec{\delta}$ connects the
    two sublattices (or the two atoms of the unit cell). (b) The electronic band structure of
    phosphorene has a band gap and is strongly anisotropic around the $\Gamma$ point. (c) An
    electrostatic potential $V$ shifts the energy bands of phosphorene and generates a pn junction,
    where electron at energy $E$ go from the conduction band (region I, orange) to the valence band
    (region II, blue).}
  \label{fig:1}
\end{figure*}

Phosphorene, a monolayer of black phosphorous, is a puckered two-dimensional crystal, see \fig{1}
(a, top). Its tight-binding model is given by
\begin{equation}
  H = \sum_{\braket{i,j}} t_{ij}\ket{i}\bra{j} + \textrm{H.c.},
  \label{Hc}
\end{equation}
where $\ket{i}$ are the atomic states localized on the phosphorous atoms at position
$\vec{r}_i$. The sum takes into account first and second nearest neighbors, which are coupled by the
energies $t_1=-1.22\un{eV}$ and $t_2=3.665 \un{eV}$, respectively \cite{Betancur-Ocampo2019,
  Rudenko, Rudenko2, Cakir, Sisakht, Ezawa, Lewenkopf, Sousa2017}. The tight-binding model of
phosphorene can be understood more easily by projecting the atoms to the $xy$ plane, keeping their
couplings constant, see \fig{1} (a, bottom). This projection shows that the phosphorene lattice can
be understood as a strongly deformed hexagonal lattice with lattice constants $a = 4.42$,
$b = 3.27 $ and $\delta=0.8 \un{\AA}$ which connects the two atoms of the unit cell \cite{Brown}.

Using a plane-wave ansatz, the tight-binding Hamiltonian can be written as
\begin{equation}
  H(\vec{k}) = \left(
    \begin{array}{@{}*{2}{c}@{}}
      0 & g^*(\vec{k})\\
      g(\vec{k}) & 0
    \end{array}
  \right),
  \label{H}
\end{equation}
where
\begin{equation}
  \label{gofk}
  g(\vec{k}) = \textrm{e}^{-\I k_{\delta}}[t_2 + 2t_1\textrm{e}^{\I k_a/2}\cos(k_b/2)]
\end{equation}
with $k_{\delta} = \vec{k}\cdot\vec{\delta}$, $k_{a} = \vec{k}\cdot\vec{a}$
$k_{b} = \vec{k}\cdot\vec{b}$. The energy bands are given by
\begin{equation}
  \label{eofk}
  \begin{array}{c}
    E(\vec{k})= s \abs{g(\vec{k})} \\[2mm]
    = s \sqrt{t_2^2 + 4 t_1 t_2 \cos(k_a/2)\cos(k_b/2) + 4 t_1^2 \cos^2(k_b/2)},
  \end{array}
\end{equation}
where $s=\text{sgn}(E)= \pm 1$. They show a band gap $2\Delta = 4t_1 + 2t_2$ and strong anisotropy
around the $\Gamma$ point. In the armchair direction ($x$-axis in \fig{1}~(a), $k_b=0$), the
electrons behave approximately as massive Dirac Fermions. Their dispersion relation is quadratic at
low energies, followed by a long linear regime. In the zigzag direction ($y$-axis, $k_a=0$), the
electrons have the parabolic dispersion of Schrödinger Fermions. Moreover, it has been shown that
this anisotropy leads to negative reflection and anti-super-Klein tunneling in phosphorene pn
junctions \cite{Betancur-Ocampo2019}. The eigenfunctions are given by
\begin{equation}
  \Psi_{\vec{k}}(\vec{r}) = \frac{1}{\sqrt{2}}\left(
    \begin{array}{@{}*{1}{c}@{}}
      1 \\
      s\textrm{e}^{\I \phi(\vec{k})}
    \end{array}\right)
  \textrm{e}^{\I \vec{k}\cdot \vec{r}},
  \label{wf}
\end{equation}
where the phase (or pseudo-spin) reads
\begin{eqnarray}
  \phi(\vec{k}) & = &
    \arctan\left(\frac{\textrm{Im}[g(\vec{k})]}{\textrm{Re}[g(\vec{k})]}\right) \nonumber\\[2mm]
                     & = &
    \arctan\left(\frac{\lambda_1(\vec{k})\cos(k_{\delta}) -\lambda_2(\vec{k})\sin(k_{\delta})}
                      {\lambda_2(\vec{k})\cos(k_{\delta}) +\lambda_1(\vec{k})\sin(k_{\delta})}\right),
   \label{phis}
\end{eqnarray}
with $\lambda_1(\vec{k}) = 2t_1\sin(k_a/2)\cos(k_b/2)$ and
$\lambda_2(\vec{k}) = t_2 + 2t_1\cos (k_a/2)\cos (k_b/2)$.

\section{Phosphorene pn and pnp junctions}


A phosphorene pn junction, which will be the main building block to construct a phosphorene
waveguide, is constituted by two regions of different electrostatic potential (or doping), which
shifts the energy bands, see \fig{1}~(c). In order to understand the electron scattering at the
interface of such pn junction, we make the following ansatz for the wavefunction in region I
\begin{equation}
  \Psi_{\text{I}}(\vec{r}) = \frac{1}{\sqrt{2}}\left(
    \begin{array}{c}
      1\\
      s\textrm{e}^{\I \phi_i}
    \end{array}\right)
  \textrm{e}^{\I \vec{k}_i\cdot\vec{r}} +
  \frac{r}{\sqrt{2}}\left(
    \begin{array}{c}
      1\\
      s\textrm{e}^{\I \phi_r}
    \end{array}\right)
  \textrm{e}^{\I \vec{k}_r\cdot\vec{r}}
\label{pwr1}
\end{equation} 
and in region II
\begin{equation}
  \Psi_{\text{II}}(\vec{r}) = \frac{t}{\sqrt{2}}\left(
    \begin{array}{c}
      1\\
      s'\textrm{e}^{\I \phi_t}
    \end{array}\right)\textrm{e}^{\I \vec{k}_t\cdot\vec{r}}.
\label{pwr2}
\end{equation}
The continuity of the wavefunction at the interface allows to calculate the
transmission coefficient
\begin{equation}
  T = \frac{2\sin[\phi_t -(\phi_r+\phi_i)/2]\sin[(\phi_i-\phi_r)/2]}{ss'-\cos(\phi_t-\phi_r)}
\label{T}
\end{equation}
as a function of the pseudo-spin angles $\phi_i$, $\phi_r$, and $\phi_t$ of the incident, reflected
and transmitted electrons \cite{Betancur-Ocampo2019}.

\begin{figure}[t]
  \centering
  \includegraphics[scale= 0.6]{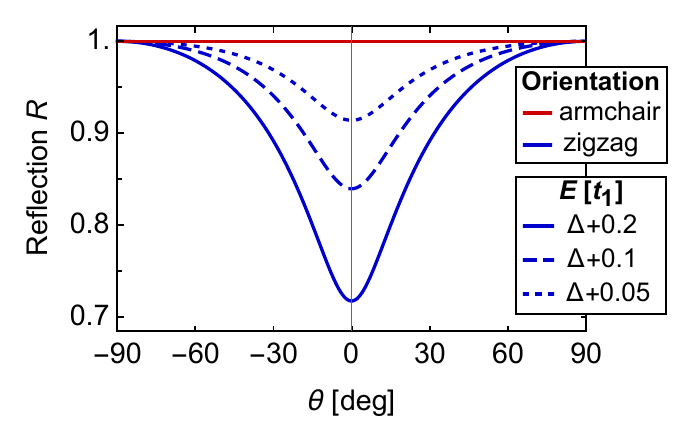}
  \caption{Reflection coefficient $R=1-T$ at the interface of a phosphorene pn junction ($V=2E$) as
    a function of the angle of incidence $\theta$ (see \fig{3}~(a)). Total reflection, $R=1$, is
    obtained for all $\theta$ and all electron energies $E$, if the junction is oriented parallel to
    the armchair edge. If the junction is parallel to the zigzag edge, the reflection is minimal for
    normal incidence but increases strongly for grazing $\theta$ and lower $E$.}
  \label{fig:2}
\end{figure}

\begin{figure}[t]
  \centering
 \includegraphics[scale= 0.4]{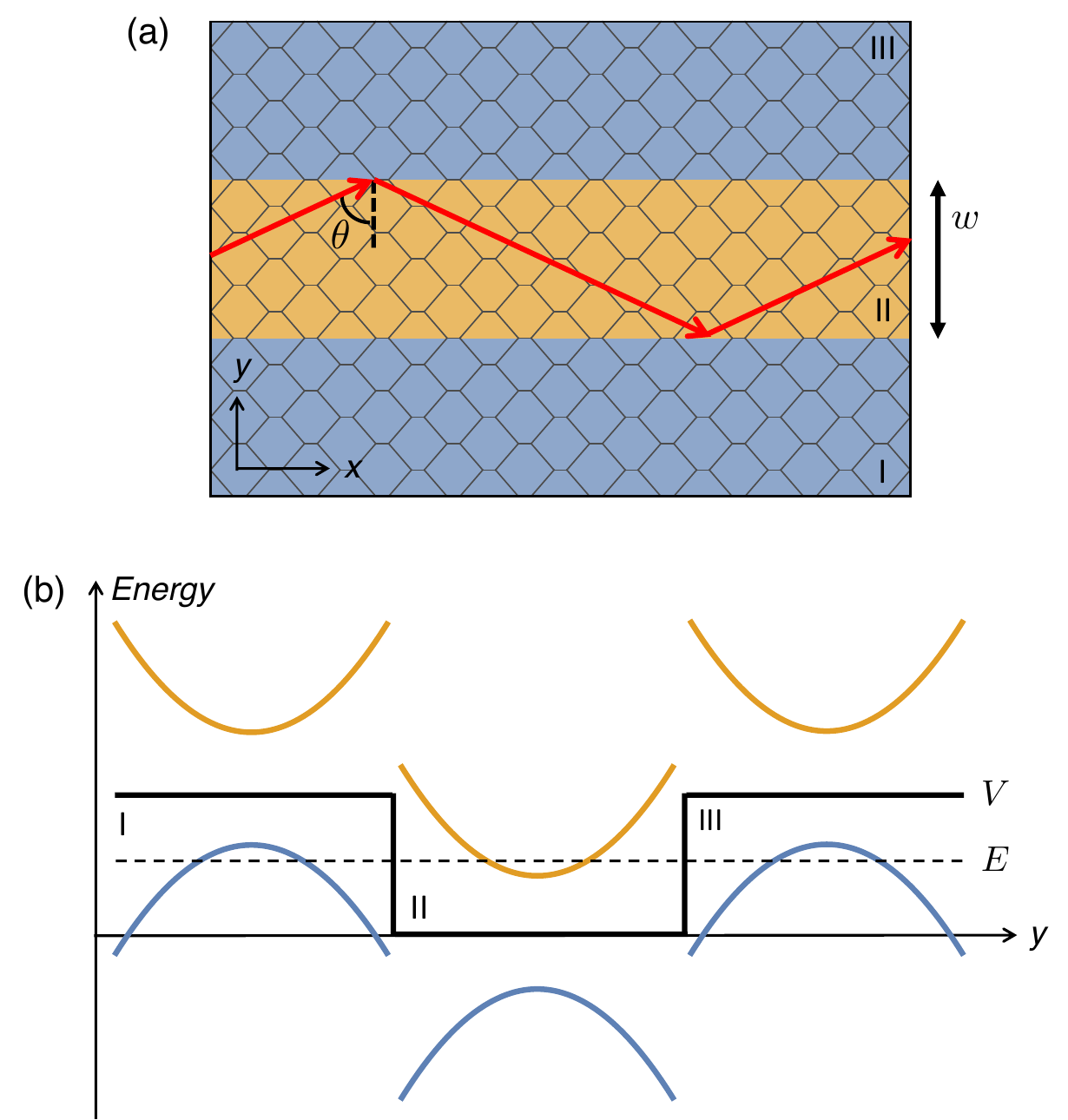}
 \caption{(a) Schematic representation of a perfect electron waveguide based on a phosphorene pnp
   junction. Electrons are injected in the n region (orange). At the interface to the p region
   (blue), they are totally reflected due to anti-super-Klein tunneling. (b) Electronic band
   structure across the potential well that generates the pnp junction.}
  \label{fig:3}
\end{figure}

\begin{figure*}[t]
  \centering
  \includegraphics[scale= 0.43]{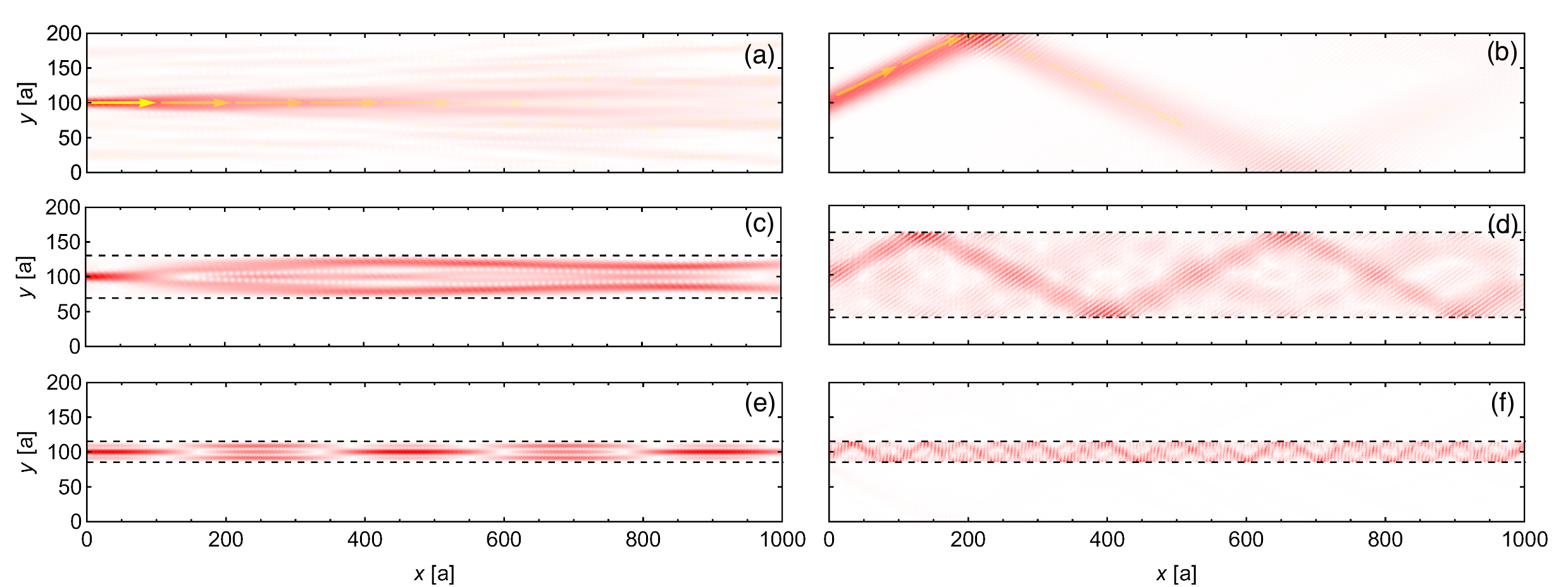}
  \caption{Current flow in phosphorene nanoribbons. The numerically calculated current density is
    indicated by the red color shading; its vector field (in some figures) by yellow arrows. The
    electrons are injected as narrow plane-waves at the left edge at energy
    $E=\Delta +0.2\abs{t_1}$. (a,b) In a pristine phosphorene nanoribbon, the ballistic ray-like
    propagation of the electrons can be observed. The electron beams are broadened due to
    diffraction and reflected at the partially absorbing edges. (c-f) In a phosphorene pnp junction,
    generated by an electrostatic potential well of hight $V= 2E$ and variable width (see the dashed
    horizontal lines), the electron beams are reflected several times within the n region but do not
    enter (not even gradually) the flanking p regions. The electrons are guided efficiently within
    the n region.}
  \label{fig:4}
\end{figure*}

Total reflection $R=1-T=1$ is obtained, if the pseudo-spins of the incident and reflected electrons
are parallel and thus, their difference vanishes, $\phi_i - \phi_r=0$. This is the case for all
incidence angles if the interface of the pn junction is aligned parallel to the armchair edge
($x$-axis in \fig{1}~(a)), because the sign change of the transverse momentum, $k_b^r= -k_b^i$, does
not affect the pseudo-spins in \eq{phis}. These findings are confirmed by \fig{2}, which shows the
reflection coefficient $R$ as a function of the angle of incidence $\theta$. This omni-directional
total reflection, named anti-super-Klein tunneling, appears for all pn junctions parallel to the
armchair edge, independently of the precise value of the electron energy and electrostatic
potential. A detailed discussion of this effect can be found in Refs. \cite{Betancur-Ocampo2019}. It
does not rely on the existence of an energetically forbidden region but on the pseudo-spin
directions. Therefore, such phosphorene pn junctions can be used to confine and guide an electron
beam within a wide range of parameters. If the pn junction is oriented parallel to the zigzag edge
($y$-axis in \fig{1}~(a)), the anti-super-Klein tunneling does not take place. Nevertheless, \fig{2}
shows that the transmission is finite and decreases for grazing angles of incidence and lower
electron energies. Waveguides constructed by means of such junction are possible but less efficient.


A perfect electron waveguide is formed by a phosphorene pnp junction that is aligned along the
armchair edge, because the anti-super-Klein tunneling confines the electrons within the n channel,
see \fig{3}~(a). Such junction is generated by the potential well
\begin{equation}
  \label{V}
  V(\vec{r})=
  \begin{cases}
    V &\text{if $ y \leq 0$ \: (sidewall, region I),}\\
    0 &\text{if  $0< y < w$ \: (channel, region II),}\\
    V &\text{if $y \geq w$ \: (sidewall, region III),}
  \end{cases}
\end{equation}
where $w$ is the width of the n channel. Experimentally, a pnp junction in phosphorene could be
realized by metallic gates. Following \cite{Cheng2019}, it could be formed also by means of a
nanotube on top of a phosphorene layer and a back-gate. Another experimental strategy could be to
deposit phosphorene (or few-layer black phosphorus) over an annealed Cu foil that produces the
potential well, as realized in graphene \cite{Bai2018}.

For the wavefunction inside the channel (region II), we make the ansatz
\begin{equation}
  \Psi_{\text{II}}(\vec{r}) = \frac{r_1}{\sqrt{2}}\left(
    \begin{array}{c}
      1\\
      s\textrm{e}^{\I \phi_i}
    \end{array}\right)
  \textrm{e}^{\I \vec{k}_i\cdot\vec{r}} +
  \frac{r_2}{\sqrt{2}}\left(
    \begin{array}{c}
      1\\
      s\textrm{e}^{\I \phi_r}
    \end{array}\right)
  \textrm{e}^{\I \vec{k}_r\cdot\vec{r}},
  \label{wfc}
\end{equation}
while outside the channel (region I/III) we write
\begin{equation}
  \Psi_{\text{I/III}}(\vec{r}) = \frac{t_\text{I/III}}{\sqrt{2}}\left(
    \begin{array}{c}
      1\\
      s'\textrm{e}^{\I \phi^\text{I/III}_t}
    \end{array}\right)\textrm{e}^{\I \vec{k}^\text{I/III}_t\cdot\vec{r}}.
\label{wfs}
\end{equation}
Using the continuity of the wavefunction at the interfaces, this ansatz yields the simple resonance
condition $\abs{k_y}w = n\pi$ with an integer number $n$. This condition can be understood easily by
taking account that anti-super-Klein tunneling takes place at the interfaces of the channel. Thus,
the sidewalls act as infinite potential walls, where the wavefunction vanishes. Note that if we
consider a potential such that the sidewalls represent energetically forbidden zones,
$\abs{V-E} \leq \Delta $, the oscillatory waves in the region I and III have to be replaced by
evanescent ones, but the resonance condition remains the same. Expanding the dispersion relation
\eqref{eofk} around the $\Gamma$ point, the quantized energy spectrum reads
\begin{equation}
  E_n(p_x) \approx \sqrt{\left(\Delta +\frac{p^2_x}{2m_x}+
      \frac{\hbar^2n^2\pi^2}{2m_yw^2} \right)^2 + v^2p^2_x},
\label{qegm}
\end{equation}
where $ m_x= 2/\left(2\delta(2at_1 - \delta\Delta)-t_1a^2\right)$ and
$m_y= -2/\left(t_1b^2\right)$ are the anisotropic masses, and
$v = at_1 - \delta\Delta$ is the velocity along the $x$ direction. The guided
modes of the waveguide are given by
\begin{equation}
\Psi_A(y) = \Psi_B(y) = \sqrt{\frac{2}{w}}\sin\left(\frac{n\pi y}{w}\right).
\label{wfgm}
\end{equation}

\section{Green's function method for quantum transport}

In order to confirm the functionality of the proposed phosphorene waveguides, we calculate
numerically the quantum coherent current flow by means of the nonequilibrium Green's function method
(NEGF) applied to the tight-binding model of phosphorene. In the following, we will summarize
briefly the essential equations. A detailed introduction to the NEGF method can be found, for
example, in Refs.~\cite{Datta, Datta2, Lewenkopf2013, Barraza-Lopez2012}.

The Green's function of the system is given by
\begin{equation}
  G(E) = (E - H - \Sigma_V - \Sigma_C)^{-1},
\end{equation}   
where $E$ is the energy of the electrons (times a unit matrix), $H$ is the tight-binding Hamiltonian
\eqref{Hc} and $\Sigma_V= \sum_n V(\vec{r}_n) \ket{n}\bra{n}$ takes into account the electrostatic
potential \eqref{V} that generates the pnp junction. In order to suppress boundary effects and mimic
an infinite system, we place a constant complex potential
$\Sigma_C= -\I \sum_{n \in \text{edge}} \ket{n}\bra{n}$ at the edges of the system, which absorbs
the electrons.

We will use two different methods to inject the electrons in the n region at the left edge, see
\fig{4}~(a). Within the first approach, the electrons are injected as plane waves with a Gaussian
profile,
\begin{equation}
  A(\vec{r}) = \exp\bigl( -\abs{\vec{r} - \vec{r}_{0}}^2/d_0^2 \bigr),  
\end{equation}
where the parameter $\vec{r}_0$ and $d_0$ control the position and width of the electron beam. These
parameters are chosen in such a way that a narrow electron beam is injected in the waveguide. The
momentum $\vec{k}$ of the plane waves, which in general does not correspond to the direction of
propagation due to the anisotropy of the electronic structure, is calculated from the input
parameters $E$ and $\theta$, as shown in Ref. \cite{Betancur-Ocampo2019}. The injection is taken
into account within the NEGF approach by the inscattering function
\begin{equation}
  \Sigma^{\text{in}}_S = \sum_{n,m \in \text{LE}}
  A(\vec{r}_n)A(\vec{r}_m) \psi^*_{\vec{k}}(\vec{r}_n)\psi_{\vec{k}}(\vec{r}_m) \ket{n}\bra{m},
\end{equation}
where the sum is over all atoms at the left edge (LE). The $\psi_{\vec{k}}(\vec{r}_m)$ are the
eigenstates in \eq{wf} evaluated at the position $\vec{r}_n$ of the atoms at the edge. Second, we
will use the wide-band model to inject the electrons. This model represents a metallic contact with
a constant surface density of states (DOS), which injects electrons with energy $E$ from the surface
of the Fermi sea without a specified direction. Within the NEGF approach, this injection is modelled
by the inscattering function
\begin{equation}
  \Sigma^{\text{in}}_S = \sum_{n,m \in \text{nLE}} \eta \ket{n}\bra{m},
\end{equation}
where $\eta \sim 1$ is a constant proportional to the DOS at the surface of the metallic
contact. The sum runs only over the atoms at the left edge which belong to the n region (nLE). The
first model has the advantage that the direction of the electron beam can be tuned precisely, which
makes it possible to compare the current flow with semi-classical trajectories \cite{Stegmann2016,
  Stegmann2019, Betancur-Ocampo2019}. The wide-band model does not have this flexibility but it is
one of the most general model that approximates various experimental situations
\cite{Verzijl2013}. Finally, the current flowing between the atoms at positions $\vec{r}_n$ and
$\vec{r}_m$ is calculated by means of
\begin{equation}
  I_{nm} = \textrm{Im}(t^*_{nm}G^{\text{in}}_{nm}),
\end{equation}
where
\begin{equation}
  G^{\text{in}} = G \Sigma^{\text{in}}_SG^{\dagger}.
\end{equation}

\section{Current flow in phosphorene waveguides}

In the following, we explore the current flow in phosphorene nanoribbons by means of the NEGF method
and compare with our predictions from the previous sections. We consider phosphorene nanoribbons of
size $(L_x,L_y)= 1000 \times 200 \,a \approx 442 \times 88 \un{nm}$, which consist of approximately
one million atoms.

In \fig{4}, we show the current flow for electrons at energy $E= \Delta +0.2\abs{t_1}$ that are
injected as narrow plane-wave beams at the left ribbon edge. Note that from now on, we measure all
energies in multiples of $\abs{t_1}$ and distances in multiples of $a$. In a pristine phosphorene
nanoribbon without an external electrostatic potential, see \fig{4}~(a,b), the electrons propagate
ballistically through the system, but the electron beams are broadened due to
diffraction. Reflections at the edges of the nanoribbon are suppressed gradually due to the
absorption of the complex potential. In presence of the electrostatic potential \eq{V}, which
generates a pnp junction, the electrons do not reach the edges at the top or bottom of the
nanoribbon but are reflected totally at the pn interfaces, see \fig{4}~(c-f). Note that the
electrons are reflected also to some degree at the right edge of the system, which leads to a
spurious counter-propagating current. The electrons are confined within the n region of the
junction, which thus forms a perfect electron waveguide. There is no leakage current, although the
electrons hit the interface under very different angles, which demonstrates the absence of critical
angles. The omni-directional total reflection at the pn interfaces, called anti-super-Klein
tunneling \cite{Betancur-Ocampo2019}, is not due to an energetically forbidden region, see
\fig{3}~(b), but due to pseudo-spin blocking. In the left column of \fig{4}, we also observe clearly
that the confinement of the electron beam in a narrow channel allows us to transfer it without
diffraction. However, note that the shown waveguides are still so wide that the quantization of the
transverse momentum $k_y$ is irrelevant and the propagation of the electron beams can be understood
in terms of ray-optics \cite{Betancur-Ocampo2019}. This perfect guiding of the current flow also
appears in asymmetric pnp' junctions as well as in npn junctions, because the anti-super-Klein
tunneling persists in these systems.

\begin{figure}[t]
  \centering
  \includegraphics[scale= 0.34]{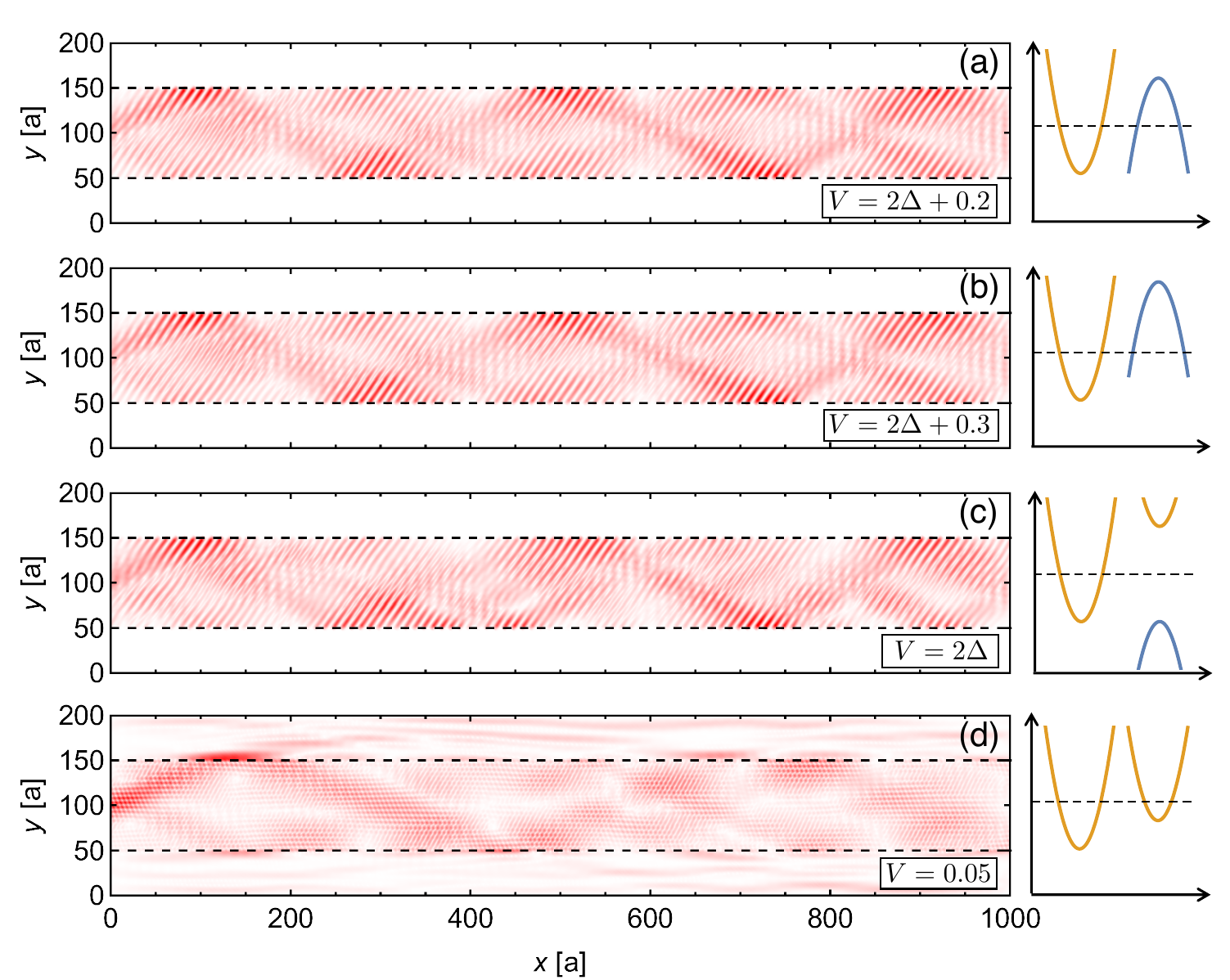}
  \caption{Current flow for electrons with energy $E=\Delta+0.1$. The energy bands in the n and p
    regions are sketched to the right. The electron waveguide does not only work for symmetric
    potential wells (a), but also for asymmetric configurations (b,c). In (c) the sidewalls are
    energetically forbidden regions due to the intrinsic band gap of phosphorene. (d) The electrons
    cannot be confined in n$'$nn$'$ junctions.}
  \label{fig:5}
  
  \vspace{6mm}
  \includegraphics[scale= 0.34]{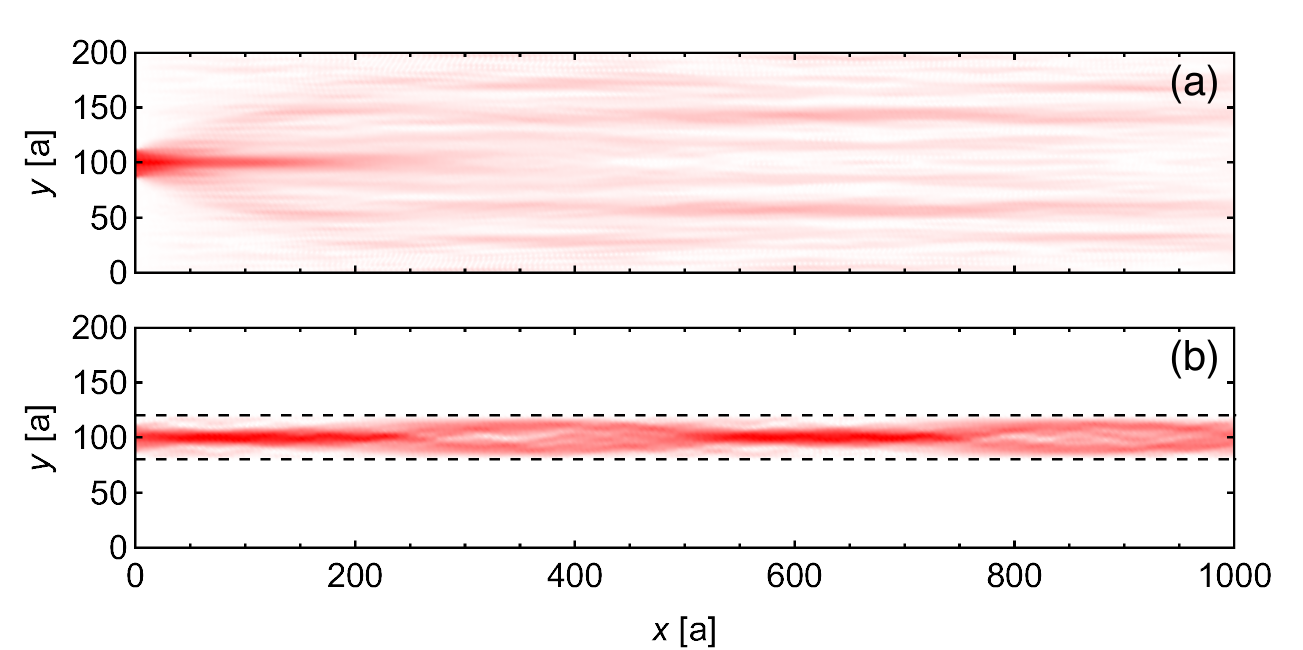}
  \caption{Current flow in the case of the wide-band model for the electron injection
    ($E=\Delta+0.1$). (a) In the homogenous nanoribbon, the current density disperses strongly due
    to diffraction. (b) A pnp junction ($V=2E$) forms a channel that guides the electrons through
    the system.}
  \label{fig:6}
\end{figure}


In \fig{5}, we demonstrate that the phosphorene waveguide also works perfectly for other electron
energies and all potentials that establish a pnp junction, $2\Delta+E < V <\infty$. Moreover, total
reflection also emerges for potentials $ E<V<2\Delta+E$ due to the intrinsic band gap of
phosphorene. As expected, the electrons cannot be confined by a n$'$nn$'$ junction.

\begin{figure}[t]
  \includegraphics[scale= 0.34]{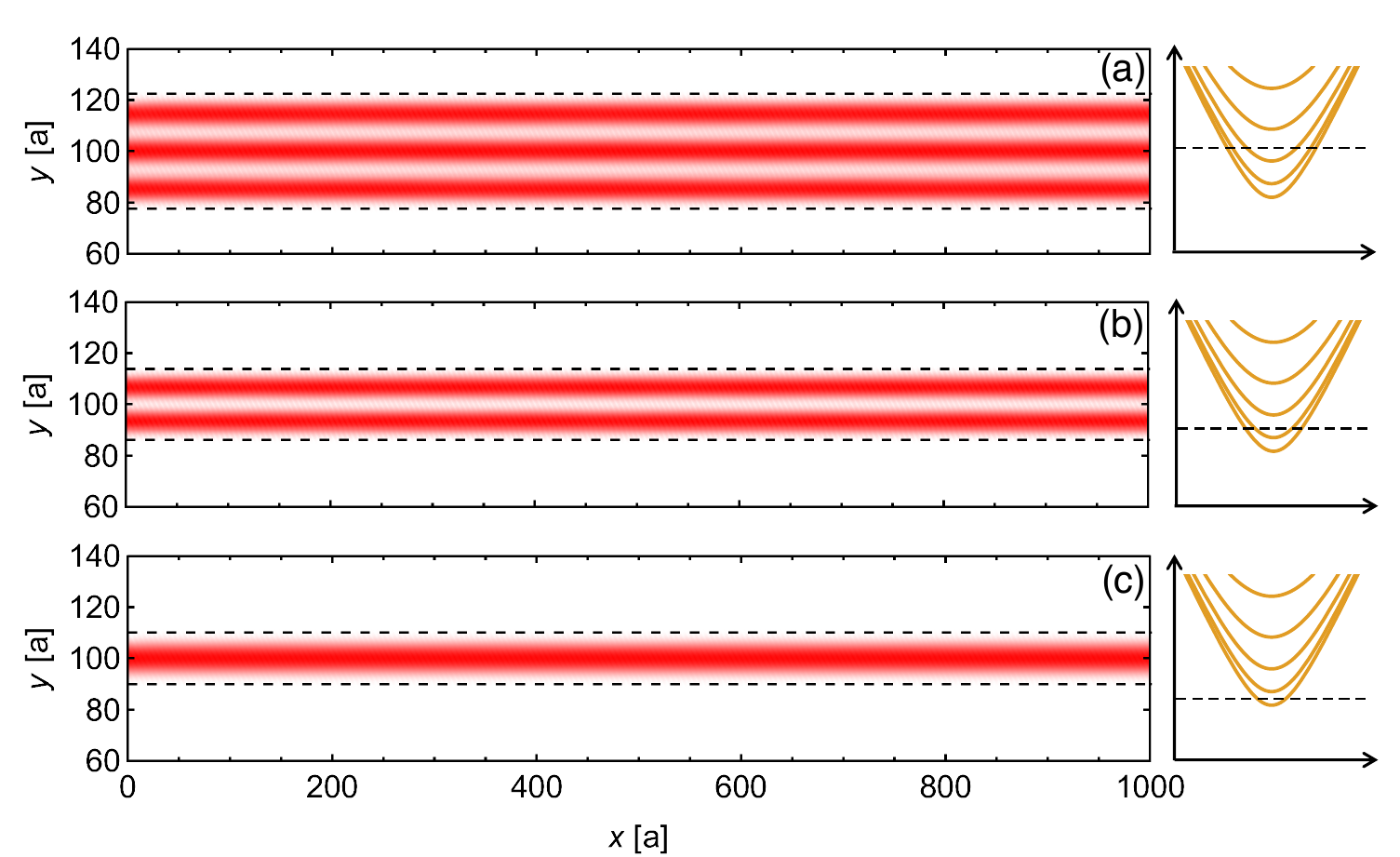}
  \caption{Current flow at low energy $E=\Delta+0.01$ in narrow waveguides. In this parameter
    regime, the subbands \eqref{qegm} due to the quantization of the transverse momentum become
    relevant and guided modes \eqref{wfgm} can be observed clearly. The subbands within the n region
    are sketched to the right.}
  \label{fig:7}
\end{figure}

In order to demonstrate the robustness of the proposed device, we show in \fig{6} the current flow
obtained by using the wide-band model for the contact that injects the electrons at the left
edge. It can be observed clearly that an otherwise dispersing electron beam is confined and guided
efficiently by the electrostatic potential of the pnp junction. The way how the electrons are
injected does not alter the functionality of the waveguide. When the electron energy and the width
of the junction is reduced significantly, the quantization of the transverse momentum $k_y$ becomes
relevant. In this regime, only some few subbands are occupied,
$n= w\sqrt{2m_y(E-\Delta)}/(\pi\hbar) \sim 1$, and the guided modes \eqref{wfgm} of the waveguide
can be observed clearly in the current flow patterns, see \fig{7}.

\begin{figure}[t]
  \centering
  \includegraphics[scale= 0.35]{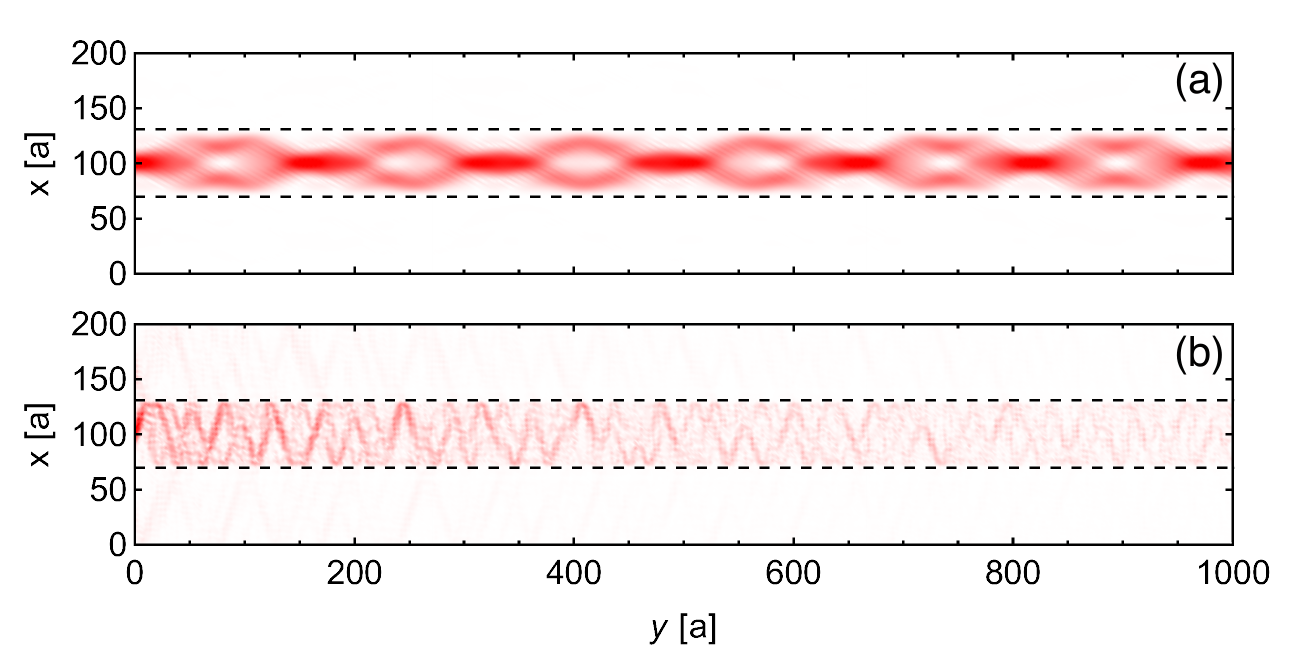}
  \caption{Current flow in a phosphorene pnp junction aligned parallel to the zigzag direction
    ($E=\Delta+0.1$, $V=2E$). The confinement of the electrons is not perfect due to the absence of
    the anti-super-Klein tunneling.  (a) At grazing incidence angles the electrons are largely
    reflected, see \fig{2}, and the junction can be used as an (albeit sub-optimal) waveguide. (b)
    At near-normal incidence, the current is leaking partially through the sidewalls. In this
    situation, the pnp junction does not serve as an electron waveguide.}
  \label{fig:8}
\end{figure}

\begin{figure}[t]
  \centering
  \includegraphics[scale= 0.35]{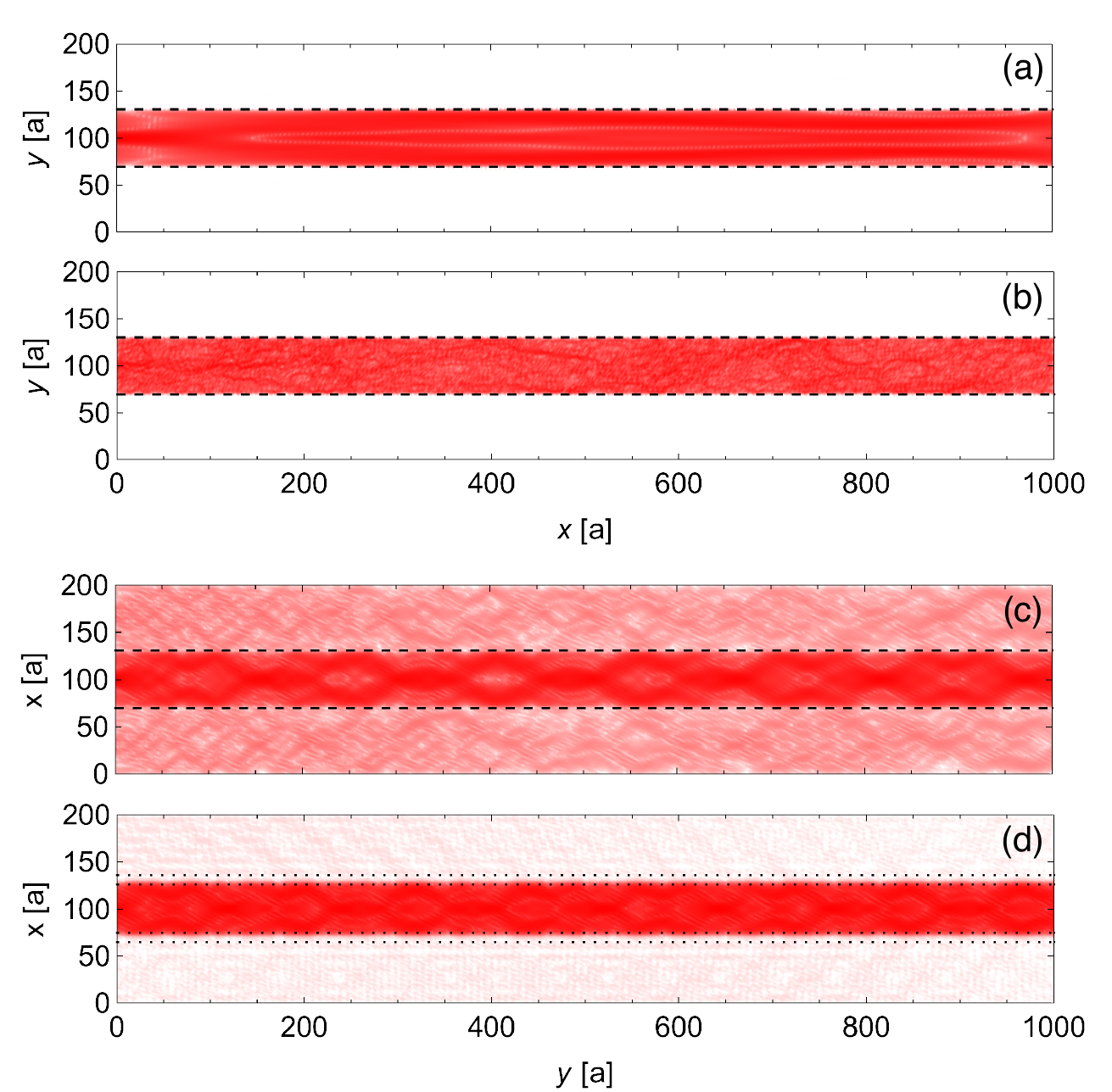}
  \caption{Current flow in pnp junctions aligned parallel to the armchair (a,b) and zigzag (c,d)
    edge, respectively. The current density is shown on a logarithmic scale. The parameters in (a)
    and (c) are the same as in Figures~\ref{fig:4}~(c) and \ref{fig:8}~(a), but the logarithmic
    scale highlights the difference between anti-super-Klein tunneling (a) and high but partial
    reflectivity (c). The parameters in (b) are the same as in (a) but we introduced random Gaussian
    disorder (standard deviation $\sigma= 0.1t_1$) and demonstrate that the anti-super-Klein
    tunneling is robust against perturbations. The parameters in (d) are the same as in (c) but the
    potential changes smoothly over 10 atomic unit cells (see the dotted lines), which induces
    energetically forbidden regions and reduces the leaking currents.}
  \label{fig:9}
\end{figure}


We have shown that the electron waveguide works perfectly, if the pnp junction is aligned parallel
to the armchair edge of the phosphorene lattice. If the channel is oriented parallel to the zigzag
edge, see \fig{8}, the confinement is not perfect because the anti-super-Klein tunneling is
absent. Nevertheless, as shown in \fig{2}, electrons that hit the interface at grazing incidence
angles are largely reflected. This effect increases even if the electron energy is lowered. Hence,
such pnp junctions constitute less efficient but applicable electron waveguides, provided that the
electrons flow essentially parallel to the sidewalls. The difference between total reflection due to
anti-super-Klein tunneling and typical high (but partial) reflectivity is shown in \fig{9}, which
gives the current density on a logarithmic color scale in order to visualize the leaking
currents. The leaking currents can be reduced significantly if the electrostatic potential changes
smoothly, because this introduces energetically forbidden regions (due to phosphorene's intrinsic
band gap) through which the electrons have to tunnel, see \fig{9}~(d). For junctions, which are not
parallel to the zigzag edge, the overall reflectivity is even higher but its minimum is obtained for
angles different from normal incidence (see Figure~4~(c) in \cite{Betancur-Ocampo2019}), which may
lead to some current leakage. However, a smooth potential could be used also in this case to
increase the confinement of the electrons. Note that we consider pristine phosphorene pnp junctions,
where the transport is ballistic. Disorder leads naturally to backscattering but still no current is
passing through sidewalls parallel to the armchair edge, because the anti-super-Klein tunneling
persists, see \fig{9}~(b).

\begin{figure}[t]
  \centering
  \includegraphics[scale= 0.34]{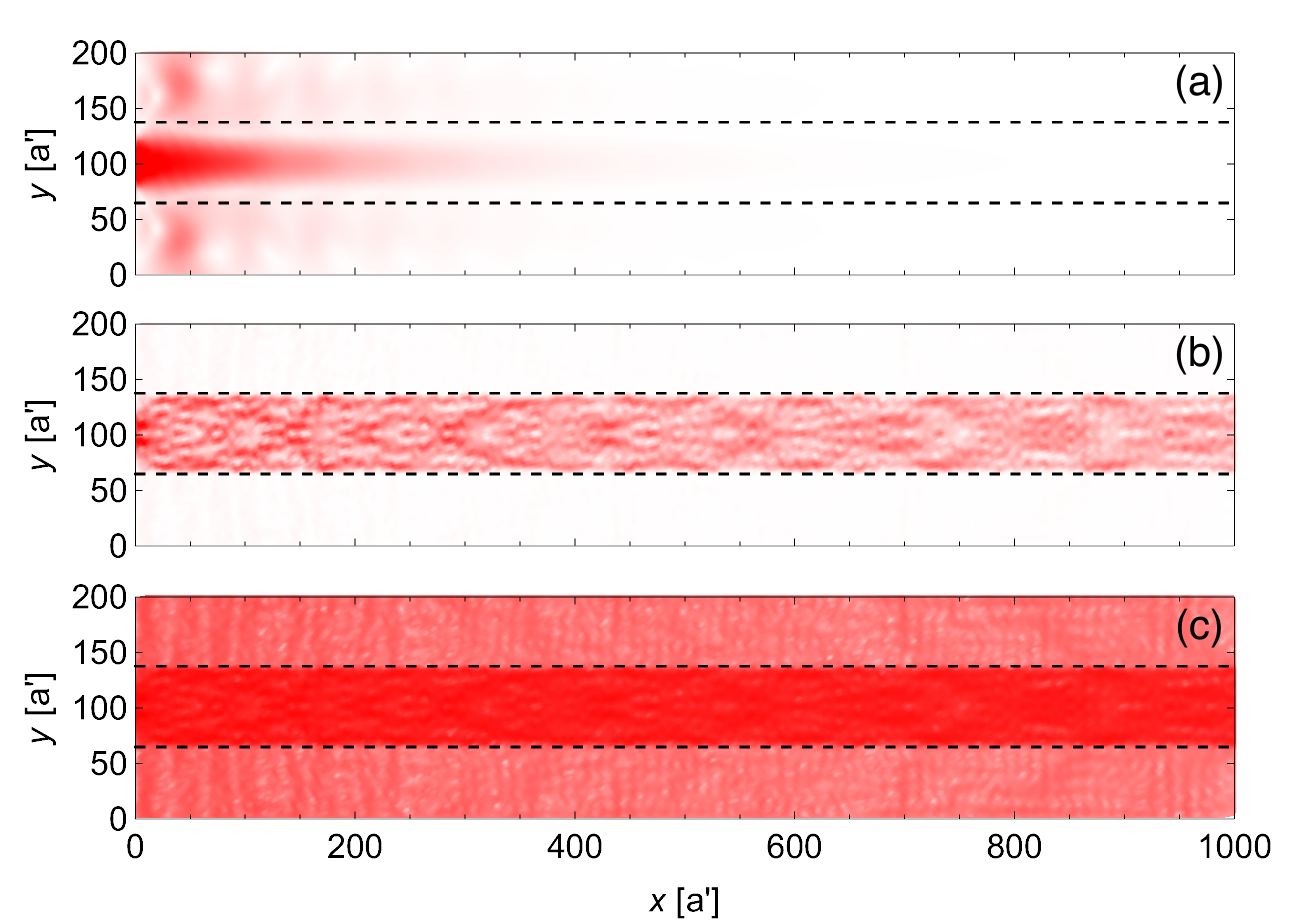}
  \caption{(a) Current flow in a graphene pnp junction ($E=0.1t_1'$, $V=2E$). Due to Klein
    tunneling, the electrons cannot be confined efficiently within the n region of the
    junction. They enter the flanking p regions and are absorbed at the top and bottom edges of the
    nanoribbon. Graphene pnp junctions cannot be used as electron waveguides. (b,c) Current flow in
    a graphene pnp junction with a staggered potential ($V_{A/B}= \pm \Delta$, $E=\Delta+0.1t_1'$,
    $V=2E$). The electrons behave like massive Dirac fermions similar to phosphorene in the armchair
    direction and are partially confined in the junction. The losses can be observed clearly on the
    logarithmic scale (c) as the anti-super-Klein tunneling is absent.}
  \label{fig:10}
\end{figure}


In order to demonstrate the unique efficiency of the proposed device, we show in \fig{10} the
current flow in a graphene pnp junction. The graphene nanoribbon is modelled by the standard
first-nearest neighbor tight-binding Hamiltonian \cite{CastroNeto2009, Stegmann2015,
  Stegmann2019}. Electrons are injected at energy $E=0.1t_1'$ by means of the wide-band model within
the n region of the junction, generated by the electrostatic potential \eq{V} with $V=2E$. Note that
in the case of graphene, energies and distances are measured in multiples of $t_1'= 2.8 \un{eV}$ and
$a'= 1.42 \un{\AA}$, the coupling energy and distance of neighboring carbon atoms. It can be
observed clearly that in graphene the electrons pass the interfaces of the pnp junction due to Klein
tunneling and are absorbed finally at the top and bottom edges of the nanoribbon. In fact, no
current density can be detected at the right edge of the system. Therefore, such a graphene pnp
junction cannot be used as an efficient electron waveguide. In \fig{10}~(b,c) we investigate the
current flow in a graphene pnp junction with broken sublattice symmetry due to the staggered
potential $V_{A/B}= \pm \Delta$ in the two sublattices A and B. In this system the electron behave
as massive Dirac fermions like in phosphorene in the armchair direction. The electrons are partially
confined in the junction but losses are observed clearly on the logarithmic scale due to the absence
of the anti-super-Klein tunneling.


\begin{figure*}[t]
  \centering
  \includegraphics[scale= 0.35]{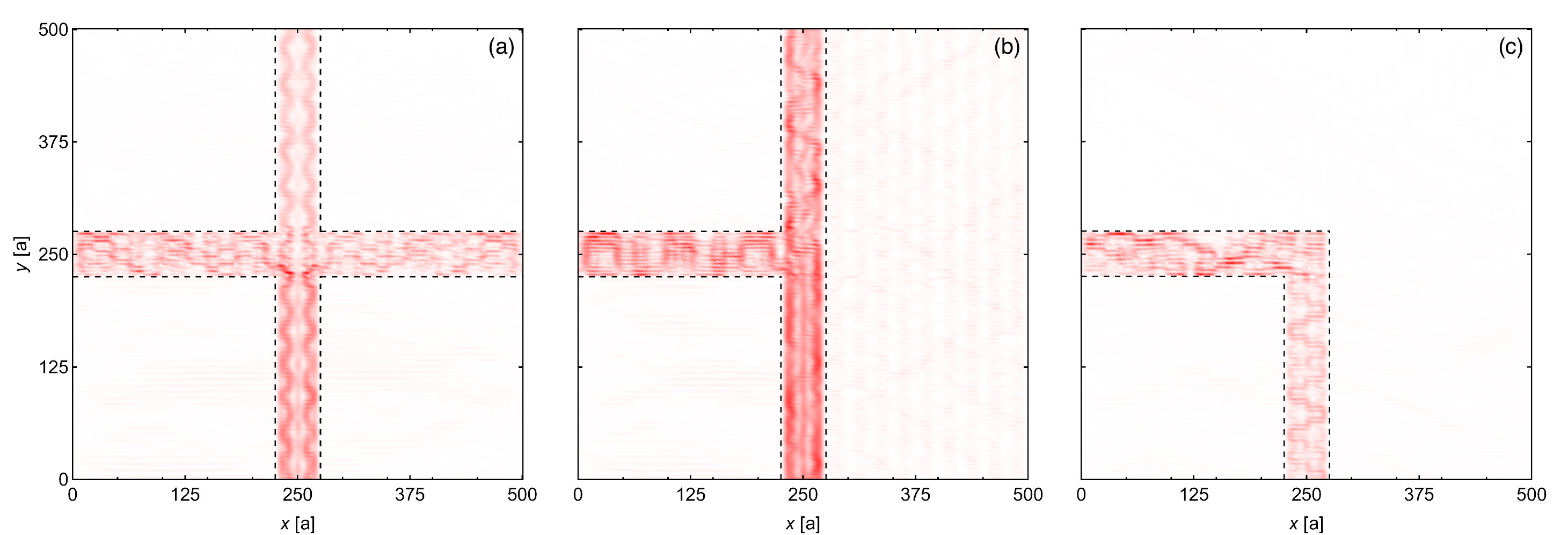}
  \caption{Current flow in a crossroad-shaped pnp junction. The current ($E=\Delta+0.1$, $V=2E$) is
    injected by means of the wide-band model at the bottom. Altough the confinement in the vertical
    waveguides is not perfect, we observe that the current is devided and guided through the n
    region of the junction. Moreover, we can block branches at the crossroad by raising the
    electrostatic potential in this region and direct efficiently the current flow.}
  \label{fig:11}
\end{figure*}

Finally, we demonstrate the current flow in a crossroad-shaped pnp junction, see \fig{11}. In order
to benefit from the high reflection at grazing incidence angles along the zigzag direction, the
current is injected at the bottom edge by means of the wide-band model. We observe that the current
is split into the different branches at the crossroad. Moreover, if branches are blocked by raising
the electrostatic potential in this region, we observe that the electron beam can be guided
efficiently in different directions. The small leakage currents of the crossroad junction will
decrease if the electron energy is reduced, because the reflectivity increases for the branches in
the zigzag direction, see \fig{2}. Note that such electron highways have been proposed also in
bilayers of graphene using the intrinsic band gap of the material \cite{Qiao2011}, but they are
impossible in monolayers of graphene.

\section{Conclusions}

We have shown that phosphorene pnp junctions, generated by regions of different electrostatic
potential, can be used to confine and guide efficiently electrons in this material. Electron beams
propagate in such waveguides like light beams in optical fibers, see \fig{4}. In narrow junctions at
low electron energy, the quantization of the transverse momentum becomes relevant and the guided
modes of the waveguide can be observed, see \fig{7}. There is absolutely no leakage current in
junctions that are aligned parallel to the armchair edge of the phosphorene lattice due to the
anti-super-Klein tunneling, the omni-directional total resistance due to pseudo-spin blocking. The
waveguides operate without any loss for all steering angles of the electron beam, because they do
not rely on the existence of a critical angle. Moreover, they work perfectly for all electron
energies and potentials that establish a pnp junction. Junctions that are not parallel to the
armchair edge suffer from partial transmission through the sidewalls of the junction, see
\fig{8}. Nevertheless, the transmission decays strongly for grazing incidence angles and lower
electron energies, see \fig{2}, making such junctions not perfect but feasible waveguides, in
particular if the steering angles of the electron beam are mostly grazing. In this way, we have
shown that a crossroad-shaped pnp junction can be used to split and direct efficiently the current
flow in phosphorene, see \fig{11}. Our theoretical work paves the way to electron optics experiments
in phosphorene. Moreover, the proposed nanoelectronic device, a perfect lossless electron waveguide,
may find also application, for example, as electron conveyors in quantum information technology.

\section*{Acknowledgments}
The authors gratefully acknowledge financial support from CONACYT Proyecto Fronteras 952, Proyecto
A1-S-13469, and the UNAM-PAPIIT research grant IA-103020. EPR gratefully acknowledges a CONACYT
graduate scholarship.  The data that support the findings of this study are available from the
corresponding author upon reasonable request.

\vspace*{5mm}

\bibliography{eofpj}

\end{document}